\documentstyle[aaspp]{article}

\newcommand{\beq}{\begin{equation}} \newcommand {\eeq}{\end{equation}}
\newcommand{\beqa}{\begin{eqnarray}}
\newcommand{\eeqa}{\end{eqnarray}} 
\newcommand{\half}{H\alpha}
\newcommand{\degr}{\,^o}
\newcommand{\teff}{T_{eff}}
\newcommand{\mvo}{M_{Vo}}
\newcommand{\ebv}{E(B-V)}

\begin{document}

\title {ARE GALAXIES OPTICALLY THIN \\TO THEIR OWN LYMAN CONTINUUM
RADIATION?\\ II. NGC 6822}
\author{Kanan Patel and Christine D. Wilson}
\affil{Department of Physics and
Astronomy,\\McMaster University,\\Hamilton, Ontario L8S 4M1 Canada}

\begin{abstract}

In this paper we study OB stars, H II regions, and the state of ionization
balance in the Local Group galaxy, NGC 6822.  Using $\half$ data and BV
photometry of the blue stars in this dIrr galaxy, we investigate the
distribution of OB stars and H II regions and determine whether individual
areas of the galaxy are separately and/or collectively in a state of
ionization balance.  Four distinct components of the $\half$ emission
(bright, halo, diffuse and field) differentiated by their surface
brightnesses are identified.  We find that approximately 1/2 of all OB stars
in NGC 6822 are located in the field while only 1/4 are found in the
combined bright and halo regions, suggesting that OB stars spend roughly 3/4
of their lifetimes outside \lq\lq classical" H II regions. If OB stars escape
from bright H II regions by destroying their parent molecular clouds, then
cloud lifetimes after forming OB stars could be as low as
$\sim$1-3$\times$10$^6$ yrs or 1/4 the typical main sequence lifetimes of OB
stars. However, if the stars are simply escaping from the clouds without
destroying them, then these data place no limits on molecular cloud
lifetimes. We find that the entire field population of OB stars cannot have
originated in and percolated out of existing H II regions.  Comparing the
observed $\half$ emission with that predicted from stellar ionizing flux
models and hydrogen recombination theory, we find that although the bright,
halo and diffuse regions are probably in ionization balance, the field
region is producing at least 6 times as much ionizing flux as is observed.
The ionization balance results in NGC 6822 suggest that star formation rates
obtained from $\half$ luminosities must underestimate the true star
formation rate in this galaxy by about 50\%.  Comparing our results for NGC
6822 with previous results for the Local Group spiral galaxy M33, we find
that the inner kiloparsec of M33 is in a more serious state of ionization
imbalance, perhaps due to its higher surface density of blue stars. Thus the
morphological class of a galaxy may be an important factor in how accurately
we can determine star formation rates from $\half$ luminosities.

\end{abstract}

\clearpage

\section{INTRODUCTION}

  Evidence that massive OB stars are not always associated with optically
luminous H II regions has been steadily increasing.  Previous indications of
this phenomenon included a study of the OB associations of M33 that showed
many of the associations do not contain bright H II regions (Wilson 1990). As
well, a study of the edge-on spiral galaxy, NGC 891, revealed the existence
of filamentary $H\alpha$ structures high above the galactic disk which are
not associated with any obvious ionizing sources (Rand et al. 1990).  In
addition, the neighbouring Large and Small Magellanic Clouds were found to
house an entire population of OB stars that were located in the \lq\lq
field" far from any OB associations (Massey et al. 1994).  These findings
raise a number of interesting question, such as how are the OB star and H II
region distributions related? What fraction of the OB star lifetime is spent
within an optically bright H II region? How prevalent are the \lq\lq field"
stars?  What is the state of ionization balance in individual and collective
regions of the galaxies?  The answers to these questions have important
implications to issues in star formation and the interstellar medium,
including the determination of extragalactic star formation rates (SFRs) and
molecular cloud lifetimes.  To answer these questions, we have investigated
the OB and H II region population and the state of ionization balance within
Local Group galaxies.  We began with a study of the spiral galaxy, M33
(Patel \& Wilson 1995, hereafter Paper I).  The present work is
focused on the dIrr member of the Local Group, NGC 6822.

We begin by outlining some of the key issues (further details can be found
in Paper I). The length of time an O star spends in the H II region phase,
and thus partially embedded in its parent cloud, will have some bearing on
the lifetime of the parent molecular cloud.  Short lifetimes ($\sim$10$^7$
yrs Blitz \& Shu 1980) assume the embedding time is a large fraction of the
total lifetime of the star so that the first O stars that form will disrupt
their parent cloud.  Molecular clouds undergoing periods of intense star
formation followed by more quiescent star formation would result in long
molecular cloud lifetimes ($\ge$10$^7$ yrs) so that each cloud could
produce several OB associations of different ages (Elmegreen 1991).  If O
stars destroy their natal environments, then the fraction of O stars seen in
the field or outside of bright H II regions (which is proportional to the
length of time a typical O star spends outside a bright H II region)
determines the time required to destroy the cloud.  However, if these field
O stars have simply moved out of their parent clouds, then cloud lifetimes
could be considerably longer.

 The second issue in this study concerns one of the most common methods of
determining SFRs in extragalactic systems, namely using the $\half$
luminosity.  This method relies heavily on the assumption that the region in
question is {\it ionization bounded} (Kennicutt 1983) i.e. all Lyman
continuum photons produced by massive young stars are absorbed by the gas in
the region.  (The alternative scenario is for a region (or galaxy) to be
{\it density bounded}, i.e. the Lyman photons are more than sufficient to
ionize all the (dense) gas.)  Since the observed $\half$ luminosity is
directly proportional to the Lyman continuum luminosity, by assuming an
initial mass function (IMF) and theoretical models for the Lyman continuum
luminosity as a function of stellar mass the \lq \lq observed" Lyman
continuum luminosity can be converted into a SFR for massive stars.  The
total SFR of a region is obtained by extending the IMF over the full range
of stellar masses.  In all of this, the assumption that the region is
ionization bounded is crucial:  since the SFR is directly proportional to
the Lyman continuum luminosity and therefore the observed $\half$
luminosity, a region out of which continuum photons are escaping will have a
smaller estimated SFR than one that is ionization bounded.  The assumption
of ionization balance is not certain to be valid if a large fraction of the
OB population lies outside bright H II regions.  In view of the wide spread
use of this technique in extragalactic systems, it is vital to verify this
important assumption of the theory.

Until very recently, ionization balance calculations in extragalactic
systems had concentrated on individual OB associations in the Large and
Small Magellanic Clouds (respectively, LMC and SMC). NGC 2122 and LH 118 in
the LMC were found to be density rather than ionization bounded (Massey
et al. 1989a) while NGC 346 in the SMC is, most likely, ionization bounded
(Massey et al. 1989b).  The results of the M33 study (Paper I) showed that
the entire (inner 1 kpc) surveyed area, as well as individual ionized gas
regions of the galaxy are not in ionization balance and in fact, in total,
the surveyed region was producing between 3 and 7 times the observed flux.

  At a distance of 0.5 Mpc (McAlary et al. 1983), NGC 6822 is very well
suited for our investigation.  While its low galactic latitude makes this a
somewhat difficult system to study, there is a substantial body of research
in the Local Group literature that is associated with the galaxy.
Photometric surveys of the stellar populations and OB associations include
the photographic surveys of Hodge (1977), and most recently, CCD imaging in
the B and V filters by Wilson (1992a).  Luminous blue stars in NGC 6822 have
been studied by Humphreys (1980), while the Wolf-Rayet star population has
been surveyed by Armandroff \& Massey (1991, 1985).  Over 140 H II regions of
the galaxy have been catalogued and studied by Hodge et al. (1989,1988).  The
recent evolutionary history, including determinations of the SFR within NGC
6822 have been discussed by Hodge et al. (1991) and Hodge (1980). Finally, the
molecular gas content of the galaxy has recently received some attention
with the CO surveys by  Wilson (1992b), Ohta et al. (1993) and Wilson (1994).

In this paper, we use $\half$ and BV photometric data of NGC 6822 to study
the distribution of the H II regions and luminous blue stars (hereafter OB
stars) in order to determine the fraction of an OB star's lifetime that is
spent outside a bright H II region and hence, to address the issues of
molecular cloud lifetimes after the formation of OB stars. With the aid of
stellar ionization models, we use the data to determine whether the
individual regions are separately and/or collectively in a state of
ionization balance.  Additionally, optical spectra of a number of stars in
NGC 6822 have been used to test the performance of the principal ionization
model used in the investigation.  The implications of the state of
ionization balance for the SFRs estimated from $\half$ luminosities are also
discussed.  Finally we compare our results for NGC 6822 with the results
obtained previously for M33. The analysis of the data ($\half$ and BV
photometry), discussion of the theoretical ionization models used and the
limitation of these models appear in section \S 2. The OB star and H II
region distributions are discussed in \S 3 and the findings of the
ionization balance calculations appear in \S 4. M33 and NGC 6822 are
compared in \S 5 and the results of our investigation are summarized in \S 6.

\section {SELECTION AND ANALYSIS  OF DATA}

\subsection{PHOTOMETRIC DATA}

We use BV photometric data from Wilson (1992a) which had previously been used
to study the OB associations in the galaxy.  The observations were made at
the Palomar 60-inch telescope using the blue sensitive Tektronix chip, CCD 6
(chip scale 0.235$^{\prime\prime}$pixel$^{-1}$).  The \lq\lq T" shaped area
surveyed included the main body of the galaxy as well as the northern bar
region which contains many OB associations previously identified by Hodge
(1977).  The average seeing in B and V was, respectively, 1.4 arcsec   and
1.3 arcsec.  Additional details of the observations can be found in Wilson
(1992a).  The average photometric uncertainties are 0.03 mag for B,V$<$20 and
0.06 mag for 20$<$B,V$<$21.  Incompleteness in the data was estimated to be
30\% for 20$<$V$<$20.5 and 55\% for 20.5$<$V$<$21.

 This data set is ideally suited for our use as it provides the most
complete (to 21 mag in V over all the surveyed regions) census of the blue
star population of NGC 6822.  Wilson (1992a) has estimated the colour excess
of the main sequence stars in the galaxy to be $\ebv$=0.45$\pm$0.05 with 0.3
mag of reddening coming from foreground sources and 0.15 mag from sources
internal to the galaxy. For the purposes of our investigation, we have
selected only those stars with V$\leq$21 and
\bv$\leq$0.5 mag.  The colour criterion, which is based on the estimated
average reddening for the galaxy, the intrinsic colour of O stars on the
zero age main sequence  ($\bv$=-0.3 mag, Flower 1977) and the observed
width of the main sequence, assures us that the maximum number of potential OB
stars will be selected for further investigation.

 \subsection[$\half$ Emission]{$\half$ EMISSION}

We used red-continuum and $\half$  CCD images of the entire galaxy
as well as its immediate surroundings kindly provided by P. Massey and G.
Jacoby.  The images were obtained using the KPNO 4-m telescope and
2048$\times$2048 T2KB chip (scale 0.48$^{\prime\prime}$ pixel$^{-1}$) on the
night of
January 7, 1992.  Exposure times were 60 sec for the red-continuum and 500
sec for the $\half$ frame. These frames have been trimmed and overscan
subtracted, flat fielded and bias subtracted in the usual manner using the
image processing package CCDPROC within IRAF.  The $\half$  image
was scaled to the same flux level as the red-continuum image using a number
of bright, unsaturated stars.  As the seeing in the $\half$ image was
slightly worse than that in the continuum image, the latter was gaussian
convolved so that stars on both images had the same full width at  half
maximum.  The final continuum-subtracted image for NGC 6822, hereafter
referred to as the $\half$ image, was then obtained by subtracting the
convolved  red-continuum image from the $\half$ image.

  The $\half$ image was calibrated by comparing published $\half$ fluxes
from Hodge et al. (1989) with the observed number of counts above the mean
background for 15 bright (surface brightness I$\ge$10$^{-13}$ erg cm$^{-2}$
sec$^{-1}$) isolated regions in the galaxy.  The slope of this relation,
which is illustrated in Figure 1, determines the uncorrected
calibration constant, $\gamma$.  As the Hodge et al. (1989) $\half$ fluxes
were measured above the atmosphere, for our purposes $\gamma$ had to be
corrected for foreground and internal extinction.  Taking
A$_{H\alpha}$=2.59$\times$\ebv=1.17 $\pm$0.13 (Schild 1977), we determined
the extinction corrected calibration constant, $\Gamma$, to be
6.5$\pm$0.8$\times$10$^{-18}$ erg cm$^{-2}$ sec$^{-1}$ count $^{-1}$. The
uncertainty in $\Gamma$ was estimated at $\sim$12\%. This figure included
the error in $\gamma$ (estimated at $\sim$2\% based on the rms error in the
slope of the $\half$ luminosity versus count graph) as well as that in
A$_{H\alpha}$.  In addition to the reddening, the data had to be corrected
for contamination from the [NII] line at 6583$\AA$ which was also observed
through the $\half$ filter used.  This was done by using the data of Pagel
et al. (1980) that give an average [NII]/$\half$ ratio of 5.5\% $\pm$1\% over
seven H II regions in NGC 6822.  Thus scaling down all counts on the $\half$
image by a factor of 0.945 corrects for the [NII] contamination in the data.

   Motived by the investigation in M33 as well as visual inspection of the
$\half$ image, we separated the galaxy into four types of emission regions:
bright, halo, diffuse and field.  The four regions were defined by their
H$\alpha$ surface brightness, {\it I}, such that {\it
I}$\geq$1.6$\times$10$^{-15}$ erg s$^{-1}$cm$^{-2}$ arcsec$^{-2}$ for the
bright regions, 4.4$\times$10$^{-16}\leq${\it I}$<$1.6$\times$10$^{-15}$ erg
s$^{-1}$ cm$^{-2}$ arcsec$^{-2}$ for the halo regions,
1.3$\times$10$^{-16}<${\it I}$\leq 4.3\times$10$^{-16}$ erg s$^{-1}$
cm$^{-2}$ arcsec$^{-2}$ for the diffuse regions and {\it I}$<$1.3
$\times$10$^{-16}$ erg s$^{-1}$ cm$^{-2}$ arcsec$^{-2}$ for the field
regions (see Figure 2).  For comparative purposes, the dividing
lines in surface brightness between the bright, halo and diffuse regions are
consistent with those adopted in M33 (without the [NII] correction applied
to the data) but that between the diffuse and field emission regions was
motived by visual inspection of the $\half$ image of NGC 6822.  We note that
the diffuse and field regions were not separated in M33 as it was not clear
whether they were two distinct components of the emission in the M33 study
or, rather, a single component that was smoothly declining in surface
brightness.  The surface brightnesses quoted above have been corrected for
extinction in $\half$ and are measured above an observed average sky surface
brightness over the galaxy of I$_{bg}$=5.32$\pm$0.53$\times$10$^{-16}$ erg
s$^{-1}$ cm$^{-2}$ arcsec$^{-2}$. The sky emission level was obtained by
averaging the modal photon counts of a number of isolated regions past the
optical limits of the galaxy.  As this emission was observed to be fairly
uniform over the entire image, it is assumed to be
\lq\lq sky" and not a smooth emission component within NGC 6822. As such, it
was not included in calculating the $\half$ luminosity of a region.  One
final point to note is that although the $\half$ emission in the entire
galaxy is available for study, we have restricted our attention to only
those regions for which BV photometric data are available i.e. the main body
and northern bar regions of NGC 6822.

\subsection[Predicted $\half$ Luminosity from OB stars]
{PREDICTED $\half$ LUMINOSITY FROM OB STARS}

 To enable a comparison of  the observed $\half$ fluxes with those
expected from the associated stellar population, we have used the stellar
ionization models of Auer \&  Mihalas (1972) (hereafter A\&M), Kurucz
(1979) and  Panagia (1973) as well as the blackbody approximation. Given the
effective temperature ($\teff$) and effective gravity ($g$) of a star, the
models predict the expected Lyman continuum flux from the star.  The
predicted ionizing flux is, in turn, used to obtain an $\half$ luminosity by
an assumed recombination scenario. The models differ in their treatment of
the stellar atmospheres. In particular, the A\&M results are based on
nonblanketed NLTE model atmospheres composed of hydrogen and helium only,
while the Kurucz models, which include the effects of line blanketing, have
LTE atmospheres of solar composition. Finally the Panagia models are based
on a combination of the NLTE A\&M models and the LTE models due to Bradley
and Morton (1969), Morton (1969) and Van Citters \& Morton (1970), and
appropriate values of $g$ are used to calculate the flux of Lyman continuum
photons as a function of $\teff$ for stars of luminosity class I, III, V as
well as the zero age main sequence (ZAMS).

We have used the calibration equations given in Parker \& Garmany (1993) to
calculate the effective temperatures and bolometric corrections (BC) for all
stars with M$_V$$\leq$21 and $\bv$$\leq$0.5.  As only B and V photometry is
available, we have used only those equations which are independent of the
reddening free index Q$\equiv$$(U-B)-0.72(B-V)$. The effective temperature
was calculated using one of equations 4b to 4e in Parker \& Garmany (1993)
while the BC of the star was determined from $\teff$ using one of equations
5a to 5d (ibid).  Effective temperatures thus calculated were used as inputs
to the ionization models while the BCs were used to estimate, where
necessary, the radius of the star emitting the ionizing flux.

 Details of the application of the various ionization models used here are
given in Paper I.  Briefly, for the  blackbody as well as
the A\&M and Kurucz models for various effective gravities, we have used
Table XIV from Massey et al. (1989b) which neatly summarizes the expected
Lyman flux at the surface of the star for $\teff$ between 20,000$\degr$ K and
50,000 K. We have adopted log($g$)=3.5 for stars with $\teff\leq$
35,500 K, and log($g$)=5.0 for stars with $\teff>$35,500 K for
the Kurucz models and log($g$)=4.0 for the A\&M models.  In order to apply
the Panagia models, we were required to determine the luminosity class of
the star as the models give the total ionizing flux (as a function of
$\teff$) separately for ZAMS, supergiant, class III and main sequence
stars.  We estimated the luminosity class of the object by comparing the
model V mag, M$_V^P$ (for a given $\teff$), with that observed. The M$_V^P$
that was closest to the observed $\mvo$ determined the luminosity class. All
stars with $\mvo\!\leq-$6.0 were taken to be supergiants (class I).

To convert Lyman continuum photons to $\half$ photons, we used the relations
N(LyC)/N($\half$) =2.22 (Osterbrock 1989) for Case B recombination, which is
appropriate for optically thick gas at 10,000 K and densities of
10$^2$-10$^4$ cm$^{-3}$, as well as N(LyC)/N($\half$)=5.36 (Brockelhurst
1971) for Case A recombination which is appropriate for gas at 10,000 K
that is optically thin in all Lyman transitions.
Clearly, Lyman continuum photons in optically thin gas produce fewer $\half$
photons than those in optically thick gas. While Case B recombination is the
most commonly used approximation in the study of H II regions, Case A
recombination was included for stars found in the diffuse and
field emission regions where the optical depth of the surrounding
gas may be smaller.

\subsection[Gauging the Reliability of the Ionization Flux]
{GAUGING THE RELIABILITY OF THE IONIZATION FLUX}

  In order to gauge the reliability of the ionization fluxes determined from
photometry plus the Panagia models, we compared the photometric spectral
classes (which determine the predicted ionizing flux due to each star) for a
number of stars in NGC 6822 with spectral classes obtained from optical
spectroscopy.  The optical spectra for 37 stars were obtained during the
nights of July 19-21, 1993 using the multi-object spectrometer ARGUS on the
CTIO 4-m telescope.  Details of the observation, reductions and spectral
classifications are described in detail
in Massey et al. (1995).  The stars selected for
spectroscopic study were some of the brightest, blue stars from the
photometric list of Wilson (1992a). Seven of the stars are found to be
Galactic foreground stars (based on their spectral classification as A or G
type) and 2 were identified as Wolf-Rayet star thus leaving only 28 OB type
stars in NGC 6822 for which we have obtained  spectral classes.
Due to the small number statistics, we have not used these stars to perform
a direct test of the ionization balance in the galaxy.

Although a number of the ARGUS spectra were strong enough to identify the
subclass as well as the luminosity class of the object, in some cases only
an approximate identification of spectral type or subclass was possible.
Spectral classes determined from photometry plus Panagia models were based
on the calculated $\teff$ and observed M$_V$ of the star (as in \S 2.3).  As
the theoretical ionizing flux models are not defined for stars of spectral
classes later than B3 (or photometrically determined $\teff$$<$$T_{eff, L}$),
all such stars were left unclassified.  The 7  Galactic foreground stars and
Wolf-Rayet stars were included in the sample in order to test the model
predictions for respectively, late type (or low $\teff$) and evolved stars.

  Before proceeding further we note that the results of this type of
comparison must be treated with caution due to the limited resolution and
signal-to-noise (S/N) of the observed spectra as well as the uncertainties
in the colour-magnitude-ionizing flux and absolute magnitude-spectral class
relations from Panagia (1973) which were used to determine the photometric
classifications.  The uncertainties in the spectroscopic classifications can
be up to 6 classes and/or up to 2 or more luminosity classes for spectra
where classification was even possible.  The reason for this large an error
is mainly due to the extreme distance of NGC 6822. Even with the long
integration times, typically 3.5 to 4 hours per star, the S/N ratio per
resolution element was 50-80. Characteristically, an S/N ratio greater than
100 is required for good identifications. To gauge the uncertainty in the
photometric spectral classification, we included the photometric errors
associated with each star to the model inputs (V and (B-V)) to find the
average change in spectral class was $\sim$4 classes.

In light of the above, we find that the match between spectral classes
determined from the ARGUS spectra with those from photometry$+$models
(uniform E(B-V)=0.45) is reasonably good: ignoring the luminosity class, of
the 37 stars, 19 or just over 50\% of the identifications agreed or were
within a few subclasses of each other.  Of the 18 stars with poor matches,
16 had photometric classifications that were significantly earlier than
those observed spectroscopically while the remaining 2 were classified later
than those observed.  The results for stars spectroscopically identified to
be earlier than O4, later than B3 or Wolf-Rayet were similar:  9 out of 17
spectral classifications were in agreement.

 In conclusion, it would appear that photometric classifications are
slightly biased towards earlier spectral types. This trend can be explained
by the presence of unresolved binary stars in the galaxy. In light of the
fact that $\teff$,  and hence the photometric classification of high mass
stars, is much  more sensitive to their V mag than their (B-V) colours (as
the colours of massive stars on the main sequence are degenerate), an
unresolved binary would be detected as a more luminous, and therefore
earlier type star then the individual stars making up the binary.
Furthermore, studies indicate that  roughly $\sim$50\% of all stars are
found in binary systems (Abt 1978; Duquennoy \& Mayor 1991) which matches
surprisingly well with the fraction of photometric identifications that did
not agree with the spectroscopic ones.  While we cannot gauge the exact
quantitative effect of the photometric bias on the theoretical ionizing
fluxes, we can safely say that its overall effect is to predict more flux
then if spectroscopic data were used.  In light of the results of this
section and for the purposes of our investigation, we shall consider any
agreement between observed and predicted fluxes that is  within a factor of
2 to be acceptable.

\section [Ionized Gas and OB Star Distribution]
 {IONIZED GAS AND OB STAR DISTRIBUTION}

  To study the common assumption that O and B stars are predominantly found
in bright H II regions, we looked at the distribution of these stars as a
function of their surrounding ionized gas environment.  We identified the
ionized gas environment of a star as either bright, halo, diffuse, or field
according to the average $\half$ surface brightness inside a circle of
radius of 3 pixels ($\sim$3.5 pc) centered on the star.  The distribution of
the OB stars (as a function of their visual magnitude)  within the four
ionized gas environments is  summarized in Table 1.

 It is immediately clear that OB stars are not all located in the bright H II
regions: in fact, $\sim$50\% of all the stars with V$<$21 and (B-V)$<$0.5
are found in the field (see Figure 3). Moreover, field stars
outnumber those in the bright regions by a factor of three.  This trend is
also suggested by the brightest stars in the survey where slightly less than
half of all stars with V$\leq$18 are seen to be in the field while only a
quarter are in the bright regions. At $\sim$11\%, the halo houses the
smallest  percentage of OB stars in the galaxy while the diffuse emission
region carries a substantial 24\% of the total.  We see faint stars dominate
all four regions and that percentage of these stars within the different
regions is comparable.

 We can now determine the important quantity, the fraction of the main
sequence lifetime of an O star that is spent within a recognizable H II
region, defined as $f_{H II}\equiv (N_B+N_H)/N_{TOT}$ where $N_{TOT}$ is the
total number of stars in the sample and $N_B$ and $N_H$ are the numbers of
stars in, respectively, the bright and halo regions of the galaxy.  By
defining a recognizable H II region to be composed of either a bright or halo
emission region, we are assured of obtaining the most conservative estimate
of this key quantity.  From Table 1,  we find $f_{H II}$=0.29,
implying that roughly 70\% of the main sequence lifetime of an OB star is
spent {\it outside} of a recognizable H II region. The consequence of this
result to the lifetimes of molecular clouds is that if OB stars escape from
bright and halo H II regions by destroying their parent molecular clouds, the
molecular cloud lifetimes after forming OB stars could be as short as
$\sim$1-3$\times$10$^6$ yrs (or 1/3 the typical main sequence lifetimes of
3-8$\times$10$^6$yrs for 120-20 M$_\odot$ stars assuming Z=.020, Schaller
et al. 1992).

  We summarize in Table 2 a few important properties of the
ionized gas environments and the associated OB star distribution.  A
striking point to note is that while the surface density of stars in the
bright region is a factor of $\sim$6 times greater than that in the field,
the field occupies an area that is $\sim$17 times that of the bright
region.  These two results provoke the puzzling question of whether the
field stars could have originated in and percolated out of existing bright
ionized gas environments.  Citing a study of the Trapezium cluster in Orion
by Zuckerman (1973), Churchwell (1991) suggests a typical O star velocity
relative to its parent molecular cloud is 3 km s$^{-1}$.  At this speed, an
O star could travel $\sim$30 pc over its lifetime of $\sim$10 Myr.  This
scales falls far short of the characteristic distances ($\sim$200-500 pc) of
the large field regions and, therefore, field stars cannot all have simply
percolated out of the H II regions currently visible.

Also from  Table 2   we find that the relative total $\half$
luminosity per field star is $\sim$4\%  that of a star in a bright
H II region,  which raises  the question of whether  the field is
\lq \lq leaking" Lyman continuum photons that presumably escape from the
galaxy. An alternative possibility is that the field population is older than
that in bright H II regions so that fewer early type stars with large
ionizing fluxes still remain. One way to investigate this possibility is to
look at the bright star distribution, which is most sensitive to age
differences, in the different ionized gas environments.  The distribution of
stars brighter than V=18, normalized to the total number brighter than V=19,
in the bright, halo, diffuse and field regions are,  respectively, 7\%
$\pm$4\%, 1\% $\pm$1\%, 5\% $\pm$3\% and 12\% $\pm$5\%. Clearly, there is no
firm indication that any one of the ionized gas environments has an
over-abundance of bright stars and therefore there is no evidence for age
segregation between the different regions.  The slope of the stellar
luminosity function can be another indicator of an age difference:  shallow
slopes imply younger stars dominate the region while steep slopes indicate
older populations dominate.  Plots of the luminosity functions, in terms of
M$_V$ versus Log(N$_*$) where N$_*$ is the total number of stars in the
magnitude bin (width 0.5 mag), are given in Figure 4.  Weighted
least-squares fits to the data to a limiting V magnitude of 19 for the
bright region (to reduce the effects of incompleteness) and V=19.5 for all
others yield slopes of 0.36 $\pm$0.05 (bright), 0.37 $\pm$0.19 (halo), 0.53
$\pm$0.16 (diffuse), 0.47 $\pm$0.07 (faint) and 0.53 $\pm$0.05 (entire
image).  Clearly there is no evidence for significant differences in the
luminosity function slopes and therefore, no indication that the stellar
populations of the four H II environments have different average ages. As
such, it appears that the differences in the $\half$ luminosity per O star
in the field compared to that in the bright regions indicates a difference
in the gas environment of the O stars and may indicate that ionizing
radiation from field stars is escaping out of the galaxy.

\section [Ionization Balance]
{IONIZATION BALANCE}

In this section, we test one of the fundamental assumptions of star
formation rate calculations from $\half$ luminosities, namely that the region
under consideration is ionization bounded i.e. the Lyman continuum flux
emitted by the stars remains in the region.  This is accomplished by
comparing the observed $\half$ emission from stars with that predicted from
stellar ionization models.  Although the detailed analysis is based on the
models of Panagia (1973), we have also used the blackbody, Kurucz, and A\&M
models for comparative purposes as well as  to serve as a check on the
Panagia results.

In all applications of the ionizing flux models, we have followed a method
similar to that described in Paper I. In brief, for comparative purposes,
we have imposed an upper and lower limit to $\teff$, $T_{eff,U}$=60,000 K and
$T_{eff, L}$=30,000 K.  The lower limit $T_{eff, L}$ is chosen as the minimum
effective temperature for which all ionization models are defined.  Although
all four models are defined up to $\teff$=50,000 K (for an O4 class V star),
we have chosen $T_{eff,U}=60,000$ K so as to provide a small margin of error
at the high $\teff$ end as well as to roughly account for the Lyman
continuum flux for stars earlier than O4. Stars that had calculated effective
temperatures between 50,000 K and 60,000 K were all assigned Lyman fluxes
corresponding to 50,000 K from the models.  Stars with $\teff <$$T_{eff,L}$
and $\teff >$$T_{eff,U}$ are assigned zero flux. Note that we have not
attempted to account for the ionizing flux from stars with effective
temperatures outside these limits:  stars with $\teff$$<$$T_{eff,L}$ are
most likely to be evolved late type stars with very small contributions to
the total Lyman continuum luminosity while those with $\teff$$>$$T_{eff,U}$
would not only entail extrapolating the $\teff$-Lyman continuum flux
relations to highly uncertain limits, but more importantly, such high
effective temperatures may be a result of photometric errors rather than
reflecting the true $\teff$ of the star (for example, Massey et al. (1989a,
1989b) have found many stars with unrealistic (B-V) or (U-B) colours which
they attributed to photometric errors). Finally, we did not consider the
presence of Wolf-Rayet stars in the  galaxy as the ionizing flux from these
stars, while likely to be very high, is rather uncertain (Massey et al.
1989b).  The effect of accounting for Wolf-Rayet stars in NGC 6822 would
be to increase the predicted ionizing flux.

  Assuming Case B recombination, the $\half$ flux from all stars with
$T_{eff,L}$$<$$\teff$
$<$$T_{eff, U}$ predicted by the four different
ionization models (in units of 10$^{39}$erg s$^{-1}$) is: 10.7 (Panagia), 7.2
(blackbody), 5.7 (Kurucz) and 6.8 (A\&M).  The highest (Panagia) and lowest
(Kurucz) estimates differ by a factor of roughly 2. Considering the
differences in the models and the arguments of \S 2.4, the agreement between
them is acceptable.  In the remainder of the section we will restrict our
discussion to the results obtained using the Panagia models only, keeping in
mind that the corresponding results from the other models scale as above.

As the minimum effective temperature for which the Panagia models are
defined depends on the luminosity class of the star, i.e. $\teff$=15100 K
(for class I), $\teff$=16,900 K (for class V and ZAMS) and $\teff$=16,000 K
(for class III), we have defined for each luminosity class  the
corresponding $T_{eff, L}$ inferred from the models.  Of the 724 stars for
which we had photometry, 428 had effective temperatures between $T_{eff, L}$
and $T_{eff, U}$. The percentage of stars with $\teff$$<$$T_{eff, L}$ or
$\teff$$>$$T_{eff, U}$ (and therefore for which no ionizing flux could be
assigned) was 58\% (V$<$18), 28\% (18$<$V$<$19), 27\% (19$<$V$<$20) and 48\%
(20$<$V$<$21).  As such, it would appear that a significant fraction of the
stars are not contributing to the total $\half$ luminosity observed.

Photometric incompleteness for all stars with 20$<$V$\leq$21 as well as
incompleteness due to stars fainter than the survey limit of 21 mag have
also been accounted for.  To correct for the photometric incompleteness of
the data, we have multiplied the predicted fluxes by a factor of 1.43 for
20$<$V$\leq$20.5 and 2.22 for 20.5$<$V$\leq$21 where the correction factors
follow from the estimated incompleteness of $\sim$30\% (for
20$<$V$\leq$20.5) and $\sim$55\% (for 20.5$<$V$\leq$21) (Wilson 1992a).  In
order to correct for stars with V$>$21, we have used a method similar to the
one used in the study of M33 (Paper I). Briefly, we calibrated a main
sequence mass-M$_V$ relation (using theoretical stellar evolution tracks for
Z=0.020 from Schaller et al. (1992)) to find that, at the distance and
reddening of NGC 6822, V=21 corresponds to 18 M$_\odot$ or roughly an O9.5
to B0 star (Popper (1980) and Schmidt-Kaler (1982)). We then used the
Salpeter initial mass function {\it N(m)d(M)=A$m^{\alpha}dm$}
($\alpha$=2.35, Salpeter 1955) to calculate the total number of stars for
each spectral (sub)class from O9.5 to B3 (the latest class for which the
Panagia models give theoretical fluxes). For each of the four ionized gas
environments, a different normalization constant $A$ was determined by
equating the integral of the initial mass function between 18 M$_\odot$ and
65.5 M$_\odot$ to the total number of stars in the region for which we could
determine $\teff$ (either uncorrected for the conservative flux estimate, or
corrected for incompleteness for the best flux estimate).  With $A$ in hand,
we then calculated the total ionizing flux in each spectral class (or
subclass) bin between O9.5 and B3 by multiply the average theoretical flux
for that bin (class V assumed) by the total predicted number of stars in the
bin.  The correction to the flux from stars fainter than V=21 was then
simply the sum of the ionizing flux from the individual bins.

  Observed and theoretical fluxes from stars in the four ionized gas
environments corrected, in turn, for the various effects described above, are
summarized in Table 3. The theoretical fluxes are organized in
the following manner: (1) no corrections for incompleteness (minimum flux
case); (2) corrected for incompleteness for stars brighter than V=21; (3)
corrected for incompleteness for V$\leq$21 and V$>$21 (best estimate) and
(4) corrected for stars below the survey limit (i.e. V$>$21).

  The best overall agreement between theory and observation is obtained for
the case where the data are
uncorrected for any incompleteness.  Over the total image, the two differ by
a factor of $\sim$1.8 (Case A used in the diffuse and field) and $\sim$2.8
(Case B used everywhere).  The predicted flux exceeds that observed by a
factor of 1.4 in the bright, 2.1 in the halo, 1.1 in the diffuse (2.8 for
Case B recombination) and 5.9 in the field (14 for Case B recombination).
While the bright, halo and diffuse regions appear to be relatively close to
ionization balance, the field is clearly not in ionization balance.
Since 50\% of the blue stars in NGC 6822 are found in the field, up to 50\%
of the ionizing photons produced may be escaping from the galaxy.

 The effect of correcting the data for V$>$21 incompleteness is minor:  the
minimum flux estimates for the four regions, separately and combined, are
increased by roughly 1\% when the flux expected from stars fainter than the
survey limits are included.  In contrast, correcting the data for
photometric incompleteness for  20$<$V$\leq$21 increases the predicted
flux from the uncorrected, uniform reddening case by a factor of $\sim$1.4.
The total theoretical flux, after this correction has been made, exceeds
that observed by a factor of 2.0 in the bright, 2.9 in the halo, 1.6 in the
diffuse (4.0 for Case B), 8.0 in the field (19 for Case B), and 2.5 for the
entire galaxy (3.8 for Case B used everywhere).  Clearly, of all the
different corrections, only that of photometric incompleteness for stars
with 20$<$V$\leq$21 is significant.  It is reassuring that the more
uncertain, model dependent correction for stars with V$>$21 is unimportant
compared to the simpler and more easily measured photometric correction.

 The best estimate of the predicted flux corrects for photometric
incompleteness over 20$<$V$\leq$21 and includes flux from main sequence
stars fainter than V=21.  Over the entire surveyed region, these fluxes are
a factor $\sim$2.5 (Case A) to $\sim$3.9  (Case B) greater than observed.
The best case for ionization balance is in the diffuse and bright emission
regions where the predicted fluxes are, respectively, $\sim$1.7 and
$\sim$2.0 (assuming Case A recombination) times those observed.  The worst
case for ionization balance is in the field, where predicted and observed
fluxes for Case A and B differ from that observed by factors of,
respectively, $\sim$8 and $\sim$20.  On average, the best estimate fluxes
are approximately 1.4  times the corresponding conservative estimates.

It is clear from Table 3  that the predicted and observed
fluxes in the diffuse and field emission regions are best matched when Case
A (optically thin) recombination is assumed.  That the gas density in these
regions should be lower than that in the bright and halo regions is not
unreasonable.  While we have no direct measure of the gas density in the
different ionization regions, we can place limits on this quantity by
comparing the observed $\half$ surface brightness with that expected for an
H II region in gas of known density.  For a B0 star
($L_{H\alpha}$=5.82$\times$10$^{35}$ erg sec$^{-1}$ and
$N_{LyC}$=4.27$\times$10$^{47}$ photons sec$^{-1}$; Panagia 1973) located in
the diffuse emission region, the maximum surface brightness observed
(I$_F$=4.32$\times$10$^{-16}$ erg s$^{-1}$cm$^{-2}$ arcsec$^{-2}$) limits
the gas density to be $\le$5 cm$^{-3}$. In obtaining the above result, we
have used the Stromgren radius (eg. $\,$Spitzer 1978) to estimate the radius
of the H II region and assumed the gas within the Stromgren sphere of the star
is completely ionized.  Similarly, if the ionizing source is an O5 star, the
gas density in the diffuse region would have to be $\le$0.5 cm$^{-3}$ for
the surface brightness of the H II region to match the maximum level
observed.  As the maximum surface density in the field is lower than that in
the diffuse region, the field gas density would be correspondingly smaller.
If the surface brightnesses of H II regions formed by B0 and O5 stars are to
equal the average surface brightness observed in bright H II regions
(I$_B$=4.11$\times$10$^{-15}$ erg s$^{-1}$cm$^{-2}$ arcsec$^{-2}$), the gas
density must be $\sim$15 cm$^{-3}$ (B0) and $\sim$1 cm$^{-3}$ (O5).  Thus
the gas density in the diffuse and field regions is likely to be lower than
that in the bright H II regions which lends some support to our assumption of
Case A recombination in the diffuse and field regions and Case B elsewhere
in the galaxy.

  What are the implications of the above findings to the determination of
high mass SFRs from $\half$ luminosities?  Recalling from our earlier
discussion that the SFR scales with the Lyman continuum luminosity, we would
expect the SFR in NGC 6822 to underestimate the \lq\lq true" SFR i.e. that
which would be measured if all the stellar ionizing radiation is converted
into, and observed as, $\half$ emission.  Over the entire galaxy, the
difference between the observed and true SFR is a factor of $\sim$2-4.
Moreover, while the difference is less than a factor of 4 in the bright,
halo and diffuse regions, in the field the discrepancy between the observed
and true SFR is a factor of $\sim$6-20.  Clearly, this technique to obtain
the rate of formation of massive stars is sensitive to the stellar
environment and cannot be used indiscriminately in any environment.  For the
field region, which houses a significant $\sim$50\% of the total number of
OB stars in the galaxy but is losing the greatest amount of ionizing flux,
this technique is most unreliable. Our results also illustrate the
importance of choosing a recombination scenario that is appropriate to the
region under investigation.

Assuming Case B recombination when Case A is more likely to be appropriate
systematically underestimates the SFR.  This may be a problem for many SFRs
that are determined using Kennicutt's (1983) relation which assumes Case B
recombination over the entire galaxy.  In NGC 6822, the SFR obtained by
applying the Kennicutt formula to the observed $\half$ luminosities in all
four ionized gas regions (SFR$_K$) underestimates by $\sim$30\% the SFR
obtained when Case A is used in the field and diffuse regions.  Moreover, by
taking the flux lost from the field region into account (by assuming the
actual $\half$ luminosity in the field is the minimum flux estimated from
the Panagia ionization models and Case A recombination in the diffuse and
field regions) we find SFR$_K$ underestimates  the true SFR by $\sim$50\%.

\subsection [Ionization Balance in Hubble I and III]
{IONIZATION BALANCE IN HUBBLE I AND III}

 What is the state of ionization balance on a small scale? To answer this
question we focus on the bright H II knot consisting of Hubble I and III
located in the northern bar of the galaxy (see Figure 5). First
identified by Hubble in 1925, Hubble I and III collectively encompass an
area roughly 0.3 kpc$^2$ (for I$\ge$1.3$\times$10$^{-16}$ erg
s$^{-1}$cm$^{-2}$ arcsec$^{-2}$).  This region was among the many H II
regions of NGC 6822 studied by Hodge \& Lee (1989) who measured the $\half$
luminosity of the region to be 6.21$\times$10$^{-16}$ erg
s$^{-1}$cm$^{-2}$.  We have summarized in Table 4  the observed
and predicted $\half$ emission in the bright, halo and combined (bright and
halo) components of Hubble I and III. The theoretical fluxes are calculated
by applying the models of Panagia (1973) to the 43 stars in the region for
which we have photometry.  In all cases, the models predict more flux than
observed in the bright emission region and less flux than observed in the
halo region.  Moreover, the agreement between theory and observations is
very good (within a factor of 1.7 to 2.3) in the bright regions but rather
poor in the halo (between a factor of 0.07 to 0.16).  This large discrepancy
in the halo region is most likely due to the very few (only 3) stars located
there.  As it turns out, the flux measured and predicted in the halo is
almost negligible when compared to that in the bright region and so does not
affect the overall conclusions of ionization balance in this individual
region.  Concentrating on the combined (bright and halo) regions, we find
that the predicted and observed fluxes are best matched when the data are
left uncorrected where the two differ by a factor of $\sim$1.6.  Correcting
for incompleteness (V$\leq$21) has the greatest overall effect on the
predicted fluxes which are seen to increase by a factor of $\sim$0.4 from
the minimum flux case and are a factor of $\sim$2.2 greater than that
observed.  Finally, the best estimate case which corrects for incompleteness
above and below V=21, yields predicted fluxes that are slightly more than
twice those observed and therefore within the allowed range of uncertainty.
Based on these results, we conclude that Hubble I and III are collectively
likely to be in ionization balance.

\section [Comparison with M33] {COMPARISON WITH M33}

  Having investigated the ionized gas-OB distribution as well as considered
the question of ionization balance in NGC 6822, we can now address the
question of how the results for this dwarf irregular system compare with
those found for the spiral galaxy M33.  In order to do so, it is important
that the colour and brightness cutoffs in the two galaxies are matched so
that the same types of OB stars are sampled.  In the case of M33, only those
stars with V$\leq$21 and (B-V)$\leq$0.4 were selected for study. Assuming a
total reddening of \ebv=0.3 mag (Wilson 1991), and a distance of 0.79 Mpc to
the galaxy (van den Bergh, 1991), the absolute magnitude cutoff
corresponding to the limit of V=21 is M$_V$=-4.4 mag.  At the NGC 6822
distance and adopted reddening of \ebv=0.45, the selection criterion for
matching the two galaxies is V$\le$20.5 and (B-V)$\leq$0.55.

  Although the diffuse emission component was not differentiated from the
field component in the M33 investigation (Paper I), we have, for sake of
comparison, done so here (see Figure 6).  The limiting surface
brightness separating the field and diffuse regions in M33, $I_{FD,
M33}$, is chosen to match that in NGC 6822 {\it before correcting for [NII]
emission} so that $I_{FD,M33}$=1.3$\times$10$^{-16}$ erg s$^{-1}$
cm$^{-2}$ arcsec$^{-2}$ uncorrected and
$I_{FD,M33}$=1.03$\times$10$^{-16}$ erg s$^{-1}$ cm$^{-2}$ arcsec$^{-2}$
corrected fro [NII] emission.  A number of interesting features of the
ionized gas distributions are revealed on comparing Figures 3  and
6.  Unlike NGC 6822,  where the diffuse emission is extended over
the galaxy, in M33 the diffuse gas appears to be confined to a narrow region
around the halo emission, thus giving the impression that the diffuse gas
is, in fact, a thin \lq\lq halo" of the latter. Additionally, the distinction
between the bright
and halo regions in M33 is much clearer than that in NGC 6822: while the
halos in M33 are broad, extended regions, those in NGC 6822 are narrow rings
or bands confined around the bright emission regions.  The field regions in
both galaxies are similar in that the emission is highly uniform and low
level.  It appears therefore that although M33 and NGC 6822 both contain
bright, halo, diffuse and field emission regions, the spatial distribution of
these components are not the same in the two galaxies. The main difference
is seen in the halo and diffuse emission regions that appear to have
exchanged roles between M33 and NGC 6822.

 In Table 5  we summarize the distribution of the
brightest OB stars ($M_V$$\le$-7.45 corresponding to V$\le$18 in M33 and
V$\le$17.5 in NGC 6822) and total number of OB stars down to the M33 survey
completeness limit V$\le$21 or $M_V$$\le$-4.45 (corresponding to V$\le$20.5
in NGC 6822).  Considering the sample of stars brighter than $M_V<$-4.45,
the results in the two galaxies are well matched in the bright and diffuse
emission components ($\sim$18\% (M33) and $\sim$17\% (NGC 6822) are located
in the bright and $\sim$24\% (M33) and $\sim$25\% (NGC 6822) in the diffuse)
but somewhat different in the halo and field regions ($\sim$32\% (M33)
compared to $\sim$11\% (NGC 6822) in the halo and $\sim$27\% (M33) compared
to $\sim$46\% (NGC 6822) in the field).  NGC 6822 has a proportionately
larger field population then M33.  Due to the significant percentage of halo
stars in  M33, the fractional lifetimes of an O star spent in a
recognizable H II region, and thus in its parent molecular cloud, is greater
in M33 ($f_{H II}\sim$50\%) than in NGC 6822 ($f_{H II}\sim$28\%).

What might the value of $f_{H II}$ be telling us about molecular clouds in
M33 and NGC 6822? If in both galaxies OB stars escape from bright and halo H
II regions by destroying their parent molecular clouds, then molecular cloud
lifetimes after the formation of OB stars are shorter in NGC 6822.  This
could be explained if the molecular clouds in NGC 6822 are smaller on
average than those in M33, since smaller clouds would be easier to destroy.
It could also be that the star formation efficiency (defined as
SFE=M$_*$/(M$_*$+M$_{gas}$) where M$_*$ and M$_{gas}$ are respectively the
mass in stars and gas in the cloud) is higher in NGC 6822 so that more stars
are formed and, therefore, the probability of destroying or dissipating the
natal molecular clouds is much increased.  Alternatively, if OB stars simply
move out of their parent molecular clouds and H II regions, then the
quantity $f_{H II}$ does not imply a molecular cloud lifetime  but suggests
that perhaps molecular clouds are smaller in NGC 6822 than M33.  Evidence
that molecular clouds may be smaller in dwarf irregular galaxies than in
spirals comes from a recent CO study of the SMC (Rubio et al. 1993).

Inspection of Table 6   reveals a number of interesting results
from the ionization balance calculations in the two galaxies.  Based on the
simple minimum flux case, the bright, halo and diffuse emission regions, as
well as the entire surveyed regions of both galaxies are likely to be
ionization bounded.  This is also suggested by the conservative and best
estimate of the ionizing flux in NGC 6822.  The problem area in the minimum
case, as in all the others, is the field which appears to be leaking the
most Lyman photons:  from $\sim$4 (in NGC 6822) and $\sim$17 (in M33) times
the observed flux in the uncorrected, minimum estimate case to $\sim$ 5 (in
NGC 6822) and a staggering $\sim$88 (in M33) times the observed flux in the
best estimate case.  The field, therefore, does not appear to be in a state
of ionization balance in either galaxy and furthermore, the problem is much
more severe in M33 than NGC 6822.  The fact that even in the minimum flux
case (where no corrections for incompleteness have been made) the field
stars in M33 are producing so much more ionizing flux than observed suggests
that the field is truly leaky to Lyman continuum photons. Although NGC 6822
has a higher percentage of field stars contributing to the ionizing flux (i.e.
stars with T$_{eff,U}<\teff<$T$_{eff, L}$ $\sim$60\% in NGC 6822 and
$\sim$45\% in M33), it is still more effective at holding on to the ionizing
photons produced by these stars than M33. This is also seen to be the case
in the other individual ionized gas environments as well as the entire
surveyed regions of the galaxies as a whole: for both the conservative and
best estimate cases, the difference between the predicted and observed
$\half$ fluxes in the bright, halo, diffuse and collective regions are
consistently higher in M33 than NGC 6822, which thus implies that the inner
kiloparsec of M33 is in a more serious state of ionization imbalance than
NGC 6822.  All three models support the finding that the likelihood of
ionization balance decreases in going from the bright to faint emission
regions. Finally, while NGC 6822 as a whole appears to be in a state of
ionization balance, the surveyed inner 1 kpc region of M33 is clearly not.

  In conclusion, it would appear that the morphological class of the galaxy
may be an important factor in these types of studies.  The fact that M33
appears to be so much less effective at holding on to the ionizing radiation
than NGC 6822 suggests that the shapes of the galaxy are important:  the
flat discs of spiral galaxies may not provide a thick enough layer of gas to
capture all or most of the Lyman continuum photons emitted by the OB stars,
especially from those stars in the field.  The surface density of stars may
also be an important factor in the extent of ionization imbalance in that
for two areas of similar column densities of neutral gas, the one with a
larger surface density of OB stars may have much less gas left to ionize and
would therefore be more likely to be leaking ionizing photons then one that
has fewer stars per unit area.  As an example consider the field regions in
M33 and NGC 6822: the M33 field star surface density is twice that of NGC
6822 ($\sim$416 stars kpc$^{-2}$ compared to $\sim$215 stars kpc$^{-2}$).
Given that the average H I column density in these two galaxies is of the
order of $\sim$10$^{20}$ cm$^{-2}$ (Deul \& van der Hulst 1987; Skillman
1987), this would mean that M33 should be twice as efficient at losing
ionizing flux as NGC 6822.  i.e.$\,$ we would expect the ratio of the
predicted to observed $\half$ luminosities in the field regions of M33 to be
a factor of two larger then that in NGC 6822. While what we observe is more
than this (in the minimum flux case, M33 is 17.09/4.07$\sim$4.2 times more
effective at losing ionizing flux than NGC 6822), the general outcome is
clear: based on the surface density of OB stars, M33 can be expected to be
more susceptible to leaking ionizing photons then NGC 6822.

\section[Conclusions] {CONCLUSIONS}

We have investigated the distribution of H II regions and OB stars and tested
the hypothesis of ionization balance within NGC 6822 using $\half$ data and
BV photometry of blue stars.  We divide the $\half$ emission in NGC 6822
into four distinct components, denoted bright, halo, diffuse and field,
based on the surface brightness of the gas.  The major findings of our study
are summarized below.

(1) The distribution of OB stars brighter than V=21 within the bright, halo,
diffuse and field regions is such that $\sim$16\% are located in the bright,
$\sim$10\% in the halo, $\sim$24\% in the diffuse and $\sim$49\% in the
field regions.  Combining the bright and halo regions reveals that only 1/4
of the blue stars are found in the optically prominent H II regions.  These
results suggest that roughly 3/4 of the main sequence lifetime of an O star
is spent outside of such  H II regions.

(2) The OB star and H II region distributions imply that if OB stars destroy
their parent molecular clouds while escaping their H II regions, then
molecular cloud lifetimes after the formation of OB stars must be shorter
than $\sim$1-3$\times$10$^6$ yrs.  Alternatively, if the stars escape the H
II regions without destroying their parent molecular clouds, then molecular
cloud lifetimes could be much longer.

(3) Comparing the spectral classes determined from photometry with those
deduced from optical spectroscopy shows that the former are biased towards
earlier spectral types, perhaps due to unresolved binaries. Due to this
effect, the theoretically predicted $\half$ fluxes are likely to be
over-estimated. We therefore consider any ionization balance calculations
that agree to within a factor of two to be in acceptable agreement.

(4) Comparing the observed $\half$ luminosities in the four ionized gas
environments both separately and collectively with those predicted
theoretically we find that the models consistently predict more flux than
observed.  The bright regions are likely to be ionization bounded as the
predicted fluxes are, at most, a factor of 2 larger than observed. The
halo and diffuse  regions appear to be at the limit of ionization balance as
the predicted fluxes can differs from that observed by up to a factor of,
respectively, $\sim$3 and $\sim$4.  In the field however, the observations
and model predictions differ by a factor of $\sim$6-15 (assuming Case A and B
recombination) for the conservative case and $\sim$8-20 for the best
estimate case. This region is clearly not in ionization balance and thus
fully 50\% of the blue stars in NGC 6822 may be losing most of their
ionizing radiation to interstellar space.

(5) In determining a SFR from the $\half$ luminosity in a galaxy, the
assumption is made that the galaxy is ionization bounded and that Case B
recombination is appropriate everywhere.  For NGC 6822, our results show
that the true SFR would be underestimated by $\sim$50\% due to the leakiness of
the field and the use of Case A recombination in the diffuse and field
regions.

(6) Significant differences between the distributions of the OB stars within
the ionized gas environments as well as the state of ionization balance in
M33 and NGC 6822 suggest that the morphological class of the parent galaxy
may be an important factor in the state of ionization balance of the gas and
therefore in determining the SFR from the $\half$ luminosity.  The field in
the inner kiloparsec of M33 appears to be in a more serious state of
ionization imbalance than NGC 6822:  M33 is losing at least $\sim$17 times
the observed $\half$ flux while the figure for the field in NGC 6822 is
$\sim$4.  The difference in the number of ionizing photon leaking from the
two galaxies may be due to the surface density of OB stars, which is higher
in M33: since the two galaxies have comparable H I column densities, the
likelihood of losing ionizing photons is increased when the stellar surface
density is increased.  Based on the ionization balance results, SFRs from
$\half$ luminosities are less likely to reflect the true SFR in the central
region of M33 than in NGC 6822.

\clearpage

{\bf FIGURE LEGENDS}

\begin{description}

\item[Figure 1.]

A graph of the observed number of counts above the mean background level
versus published $\half$ fluxes  (in units of 10$^{-18}$ erg cm$^{-2}$
s$^{-1}$) from Hodge et al. (1989) for 15 bright (surface brightness
I$\ge$10$^{-13}$ erg cm$^{-2}$ sec$^{-1}$) isolated regions in the
galaxy.   The best fit slope through the data determines the
calibration constant $\gamma$=2.22$\times$10$^{-18}$
erg cm$^{-2}$ s$^{-1}$ count$^{-1}$.

\item[Figure 2.]

The $\sim$7.5$^\prime\times$7.5$^\prime$ continuum subtracted $\half$ image of
NGC
6822 illustrating the distribution of the bright, halo, diffuse and field
$\half$ emission regions.  North is at the top and east is to the left.  For
reference, the giant H II regions Hubble I and III are to the top right.  The
contours, corresponding to surface brightnesses of {\it
I}=1.7$\times$10$^{-15}$, 4.6$\times$10$^{-16}$ and 1.4$\times$10$^{-16}$
erg s$^{-1}$cm$^{-2}$arcsec$^{-2}$, separate the bright, halo, diffuse and
field regions.

\item[Figure 3.]

The distribution of all OB stars (V$\leq$21 and (B-V)$\leq$0.5) in the inner
kpc of NGC 6822.  Photometric coverage of the galaxy was restricted to
inside the \lq\lq T" shaped region outlined. The location of field OB stars
are marked by plus signs while the  OB stars in the bright, halo and diffuse
regions are marked by a small square.

\item[Figure 4.]

The luminosity functions for all blue ((B-V)$\leq$0.5) stars in the survey as
well those in the bright, halo, diffuse and field ionized gas environments.
An arbitrary offset in the ordinate has been applied for the purposes of
clarity. Overlaid on the data are the weighted least squares fits (to a
limiting magnitude of V=19 for the bright region and V=19.5 elsewhere).

\item[Figure 5.]

The $\half$ image of the giant H II regions, Hubble I and III. Overlaid are
the OB stars within the bright and halo emission regions

\item[Figure 6.]

The continuum subtracted $\half$ image of the central
8.7$^\prime\times$8.7$^{\prime}$ of M33 illustrating the distribution of the
bright, halo, diffuse and field $\half$ emission regions.  North is at the
top and east is to the left. For orientation the giant H II region NGC 595
is located in the top right-hand corner of the image and the center of the
galaxy coincides roughly with the center of the image.

\end{description}
\clearpage

\begin{table*}
\begin{center}
\begin{tabular}{lcccccccccc}
\tableline
\tableline
\multicolumn{1}{c}{
V MAG   \tablenotemark{*}}
&  \multispan2\hfil  BRIGHT  \hfil
&  \multispan2\hfil  HALO  \hfil
&  \multispan2\hfil DIFFUSE   \hfil
&  \multispan2\hfil FIELD  \hfil
&\multispan2\hfil   ENTIRE IMAGE\hfil\\
&   \# stars & \% region &   \#  stars & \%  region
&   \#  stars & \%  region & \#  stars & \%   region
& \# stars & \%  region \\
\tableline
\tableline
V  $<$18 &  5  $\pm$2  &  4 $\pm$2 \%  &  1 $\pm$1  &  2 $\pm$2 \%
	 &  4  $\pm$2  &   2$\pm$1 \%  &  9 $\pm$3  &  3$\pm$1 \%
	 &  19  $\pm$4  &   3$\pm$1 \%  \\
18$<$V$<$19
	 &  16   $\pm$4  &  13$\pm$4\%  & 8  $\pm$3  &  10$\pm$5\%
	 &  6 $\pm$2 &   3$\pm$1 \%  & 25 $\pm$5  &  7$\pm$2 \%
	 &  55  $\pm$7  &   8$\pm$1 \%  \\
19$<$V$<$20
	 &  35  $\pm$6  &  30$\pm$8 \%  & 27  $\pm$5  &  35$\pm$11\%
	 &  44   $\pm$7	&   25$\pm$6 \% & 88 $\pm$9   &  25$\pm$4 \%
	 &  194  $\pm$14  &  27$\pm$3 \%  \\
20$<$V$<$21
	 &  62   $\pm$8 &  52$\pm$11\%  & 41  $\pm$6 &  53$\pm$14\%
	 &  121  $\pm$11&  69$\pm$11 \% & 232 $\pm$15 &  67$\pm$8 \%
	 &  456  $\pm$21  &   63$\pm$5 \%  \\
\tableline
V$\leq$21(Total)& 118 $\pm$11 &  &  77 $\pm$8 & & 175 $\pm$13 & & 354 $\pm$19
& & 724 $\pm$27  \\
\tableline
\tableline
\end{tabular}
\end{center}

\tablenotetext{*}{Note : Only stars with (B-V)$\leq$0.5}
\caption{The Distribution of Blue  Stars in Four Ionized Gas Environments in
N6822.}
\label{tbl-numregion}
\end{table*}

\clearpage
\begin{table*}
\begin{center}
\begin{tabular}{lccccc}
\tableline
\tableline
&  BRIGHT & HALO & DIFFUSE &FIELD & ENTIRE IMAGE \\
\tableline
\tableline
\multicolumn{1}{l}{Total \# of Stars \tablenotemark{a}}
& 118  & 77  & 175  & 354 & 724\\
Area Covered  (kpc$^2$)  & 0.06  & 0.13  &  0.56  & 1.04 & 1.79 \\
Surface Density of Stars (kpc$^{-2}$)
&  1967 & 592 & 312 & 340 & 404 \\
\multicolumn{1}{l}{Average Surface Brightness [I] \tablenotemark{b}} &
71.56 & 8.03 & 0.99 & 0.20 & 1.44  \\
\multicolumn{1}{l}{Total Luminosity [L$_{ob}$]  \tablenotemark{c}}
& 2.19 & 0.53 & 0.83 & 0.29 & 3.85 \\
\multicolumn{1}{l}{Relative  Luminosity per Star \tablenotemark{d}} &
1 & 0.37 & 0.26 & 0.04 & 0.28 \\
\tableline
\tableline
\end{tabular}
\end{center}
\caption{Average Properties of the Four Ionized Gas Environments.}
\label{tab-ch3-2}
\end{table*}


$^{a}$ All stars  with (B-V)$\leq$0.5 and V$\leq$21.

$^{b}$ In units of 10$^{-16}$ erg s$^{-1}$cm$^{-2}$
arcsec$^{-2}$.This is measured above the average background.

$^{c}$ In units of L$_{39}\equiv$10$^{39}$
erg s$^{-1}$.

$^{d}$ In units of 1.86$\times$10$^{37}$ erg s$^{-1}$
star$^{-1}$.

\clearpage

\begin{table*}
\begin{tabular}{lccccc}
\tableline
\tableline & BRIGHT & HALO & DIFFUSE & FIELD & TOTAL \\
 & & & (CASE A/B) &(CASE A/B) & (CASE A/B) \\
\tableline
\multicolumn{1}{l}{Minimum Flux\tablenotemark{a}}
 &3.16 & 1.10 & 0.95 / 2.29 & 1.72 / 4.16 & 6.93 / 10.71  \\
\multicolumn{1}{l}{Incompleteness (V$<$21)\tablenotemark{b}}
& 4.28 & 1.52 & 1.36 / 3.28 & 2.33 / 5.64 & 9.75 / 14.72\\
\tableline
\multicolumn{1}{l}{Best Estimate \tablenotemark{c}}
 &  4.32 & 1.55 & 1.39 / 3.33 & 2.39 / 5.73 & 9.65 / 14.93 \\
\multicolumn{1}{l}{Conservative  Estimate \tablenotemark{d}}
& 3.19 & 1.22 & 0.96 / 2.32 & 1.74 / 4.22 & 7.11 / 10.95 \\
\tableline
\multicolumn{1}{l}{Observed Flux\tablenotemark{e}}
& 2.19 & 0.53 & 0.83 & 0.29 & 3.85  \\
\tableline
\tableline
\end{tabular}


\caption {Observed and Predicted  H$_\alpha$ Flux
in  the  Four   Ionized Gas  Environments.}
\label{tab-ch3-3}

\end{table*}

$^{*}$ Note: All stars with V$\leq$21 and
all  fluxes in  units of L$_{39}$$\equiv$10$^{39}$erg s$^{-1}$.

$^{a}$ Minimum: no corrections applied.

$^{b}$ Incompleteness: corrected
for  incompleteness in  stars brighter than V=21.

$^{c}$ Best estimate: corrected for
stars   brighter and fainter than  V=21.

$^{d}$ Conservative estimate:  corrected
only for stars with  V$>$21.

$^{e}$  Observed fluxes are measured above the  average sky
background.

\clearpage

\begin{table*}
\begin{tabular}{lccc}
\tableline
\tableline
                &  BRIGHT       & HALO          & COMBINED \\
                &               &               & (BRIGHT $+$ HALO)\\
\tableline
\multicolumn{1}{l}{Minimum Flux\tablenotemark{a}}
& 1.24 &  0.0035 & 1.24 \\
\multicolumn{1}{l}{Incompleteness (V$\le$21)\tablenotemark{a}}
& 1.72 & 0.0080 & 1.73  \\
\tableline
\multicolumn{1}{l}{Best Estimate \tablenotemark{a}}
& 1.74  & 0.0082 & 1.73 \\
\multicolumn{1}{l}{Conservative  Estimate \tablenotemark{a}}
&  1.25 & 0.0049 &  1.26 \\
\tableline
\multicolumn{1}{l}{Observed Luminosity [L$_{ob}$]\tablenotemark{b}}
& 0.75 & 0.05  & 0.80  \\
\# Stars Observed [$N_{ob}$]  & 40 & 3 & 43 \\
\multicolumn{1}{l}{\# Stars with $T_{eff, L}<\!t_{eff}<T_{eff, U}$
\tablenotemark{c}}  & 35 & 3 & 38  \\
\multicolumn{1}{l}{Relative Luminosity per star [L$_{ob}$/$N_{ob}]
$\tablenotemark{c}}
&1.01 & 0.89 & 1.00 \\
\tableline
\tableline
\end{tabular}

\caption{The Observed and Predicted H$\alpha$ Fluxes
and OB Star  Content in Hubble I and III.}
\label{tab-ch3-5}
\end{table*}

$^{a}$ Predicted $\half$ luminosity accounting for incompleteness
and/or differential reddening (as in Table 3).  All stars with
V$\le$21 and all fluxes in units of L$_{39}$.

$^{b}$ Observed luminosities have been  background subtracted.

$^{c}$ In units of 1.86$\times$10$^{37}$erg s$^{-1}$ arcsec$^{-2}$
star$^{-1}$.

\clearpage

\begin{table*}
\begin{tabular}{llccccc}
\tableline
\tableline
                & GALAXY & BRIGHT & HALO & DIFFUSE & FIELD & TOTAL \\ & & \#
of stars &\# of stars &\# of stars &\# of stars &\# of stars \\
\tableline
\multicolumn{1}{l}{
Brightest Stars    \tablenotemark{a}}
&  M33  & 22 $\pm$5     &  15  $\pm$4   & 7 $\pm$3& 11 $\pm$3.3 & 55  $\pm$7
\\
&NGC 6822&  2 $\pm$1    & 0     & 2 $\pm$1      & 6 $\pm$2 & 10 $\pm$3 \\
&       &               &               &       &    &             \\
\multicolumn{1}{l}{
Total in Survey    \tablenotemark{b}}
&  M33  &390 $\pm$20    & 688 $\pm$26   &519 $\pm$23 & 583 $\pm$24& 2180
$\pm$47 \\
&NGC 6822&  84 $\pm$9   & 51 $\pm$7     & 119$\pm$11    & 224 $\pm$15 &  481
$\pm$22 \\
\tableline
\tableline
\end{tabular}



\caption[Distribution of OB Stars in M33 and NGC 6822] {The Distribution of
OB Stars in M33 and NGC 6822.}

\end{table*}

$^{a}$  The brightest stars
are selected such that  $M_V<$-7.45  in both galaxies and
(B-V)$<$0.4  for M33 and (B-V)$\le$0.5 for NGC 6822.
Note  that $M_V\le$-7.45  corresponds to
V$\le$18 in M33 and V$\le$17.5 in NGC 6822.

$^{b}$  Total number of stars to a limiting absolute
magnitude $M_V\le$-4.45
(corresponding to  V$\le$21  in M33 and V$\le$20 in NGC 6822)
for  both galaxies. The colour cutoff as is above.

\clearpage

\begin{table*}
\begin{center}
\begin{tabular}{llccccc}
\tableline
\tableline
& GALAXY& BRIGHT& HALO & DIFFUSE & FIELD & TOTAL \\
&        & (CASE B)& (CASE B)& (CASE A)  & (CASE A) &  \\
\tableline
MINIMUM
& \multicolumn{1}{l}{M33\tablenotemark{a}}
& 1.29 & 1.58 & 1.30 & 17.09 & 2.29  \\
& \multicolumn{1}{l}{NGC 6822\tablenotemark{b}}
& 1.14 & 1.54 & 0.91 & 4.07 & 1.46 \\
& & & & \\
CONSERVATIVE
& \multicolumn{1}{l}{M33\tablenotemark{c}}
& 2.10 & 3.62 & 2.98 & 40.11 & 5.28    \\
& \multicolumn{1}{c}{NGC 6822\tablenotemark{d}}
& 1.20 & 1.74 & 1.00 & 4.66 & 1.49  \\
& & & & \\
BEST ESTIMATE
& \multicolumn{1}{l}{M33\tablenotemark{e}}
 & 3.97 & 8.35 & 6.20 & 88  & 6.51 \\
& \multicolumn{1}{l}{NGC 6822\tablenotemark{f}}
& 1.29 & 1.73 & 1.13 & 4.93 & 1.59 \\
\tableline
 \multicolumn{1}{l}{OBSERVED H$\alpha$ \tablenotemark{g}}
& M33     & 3.78 & 2.57 & 0.95  & 0.06  & 7.36 \\
& NGC 6822& 2.19 & 0.53 & 0.83 & 0.29 & 3.85 \\
\tableline
\tableline
\end{tabular}
\end{center}









\caption{Predicted and Observed H$\alpha$ Fluxes in the
Four   Ionized Gas Environments of M33 and NGC 6822.}

\end{table*}

$^{ }$ All predicted fluxes are expressed as a fraction of the
observed $\half$ flux in the particular  ionized gas environment of the
galaxy.  Magnitude and colour cutoffs in respectively, M33 and NGC 6822 are,
respectively, V$\le$21 and (B-V)$\le$0.4 and V$\le$20.5 and
(B-V)$\le$0.55.

$^{a}$ Minimum flux M33:   no corrections
applied.

$^{b}$ Minimum flux NGC 6822: no
corrections applied.

$^{c}$ Conservative estimate M33:
corrected only for stars with V$>$21.

$^{d}$ Conservative estimate NGC 6822:
corrected only for stars with  V$>$20.5.

$^{e}$ Best estimate M33: corrected for
stars with  V$\leq$21 and V$>$21.

$^{f}$ Best estimate NGC 6822:  corrected for
stars with  V$\leq$20.5 and V$>$20.5.

$^{g}$ Observed fluxes are in units of L$_{39}\equiv$10$^{39}$
erg s$^{-1}$ arcsec$^{-2}$.

\end{document}